\documentclass[oneside, reqno,11pt]{amsart}
\usepackage{graphicx}
\usepackage{caption, subcaption, float}
\usepackage{bigints,array}
\usepackage[margin=1in]{geometry}
\usepackage{amsmath}

\newtheorem{theorem}{Theorem}[section]

\theoremstyle{definition}

\newtheorem{conj}[theorem]{Conjecture}

\theoremstyle{remark}

\numberwithin{equation}{section}

\begin{document}

\title{Map enumeration from a dynamical perspective}

\author{Nicholas Ercolani}
\address{University of Arizona}
\curraddr{}
\email{ercolani@math.arizona.edu}
\thanks{}

\author{Joceline Lega}
\address{University of Arizona}
\curraddr{}
\email{lega@math.arizona.edu}
\thanks{}

\author{Brandon Tippings}
\address{University of Arizona}
\curraddr{}
\email{tippings@arizona.edu}
\thanks{}

% ORCID Info:
% Nick - 0000-0003-2010-4205
% Joceline - 0000-0003-2064-229X
% Brandon - 0000-0001-7309-7040


\date{}

\begin{abstract}
This contribution summarizes recent work of the authors that combines methods from dynamical systems theory (discrete Painlevé equations) and asymptotic analysis of orthogonal polynomial recurrences, to address long-standing questions in map enumeration. Given a genus $g$, we present a framework that provides the generating function for the number of maps that can be realized on a surface of that genus. In the case of 4-valent maps, our methodology leads to explicit expressions for map counts. For general even or mixed valence, the number of vertices of the map specifies the relevant order of the derivatives of the generating function that needs to be considered. Beyond summarizing our own results, we provide context for the program highlighted in this article through a brief review of the literature describing advances in map enumeration. In addition, we discuss open problems and challenges related to this fascinating area of research that stands at the intersection of statistical physics, random matrices, orthogonal polynomials, and discrete dynamical systems theory.
\end{abstract}

\maketitle

\section{Introduction: map enumeration, random matrices, and orthogonal polynomials}
\label{sec:intro}

The topic of cellular decomposition of surfaces, including the fundamental idea of triangulation in algebraic topology, extends back to the origins of combinatorial topology. To this day, a multitude of applications of this concept and associated enumerative questions have arisen in the physical, biological, and social sciences \cite{bib:ba99,bib:cn06,bib:le11,bib:fomg13,bib:mb14,bib:hs21,bib:fr23}.

A systematic analysis of the classification and enumeration of such decompositions arose only about 60 years ago in the pioneering work of Tutte and his collaborators \cite{bib:tu68}. Tutte's original focus was on enumerating cellular decompositions of the plane (equivalently the sphere) with a fixed number of vertices. He referred to this as \emph{map enumeration} due to the evident analogy with the 4-color problem for geographical maps. It is clear in the case of the plane or sphere that map enumeration can be directly related to the problem of enumerating graphs constrained to lie on the plane. Tutte formulated a program combining methods of low-dimensional topology and classical graphical enumeration to come up with a recursive algorithm for enumerating \emph{planar maps}. In that context, a map is simply a connected graph embedded in a surface, whose complement tessellates the surface, possibly including a tile at infinity. Subsequent work by Brown \cite{bib:br66} and Arqu\`es \cite{bib:ar87} opened the door to studying map enumeration for low genus surfaces. (Passing to genus $g >0$ entails further topological constraints that will be described below.) Bender and Canfeld \cite{bib:bc86} extended Tutte's equations to surfaces of general genus and initiated a program of asymptotic (in vertex number) enumeration. More recently Eynard \cite{bib:ey11,bib:ey16} obtained a refined version of their recursion. With all these developments, map enumeration became established as a discipline in its own right.

\begin{figure}[ht]
\includegraphics[width=.6 \linewidth]{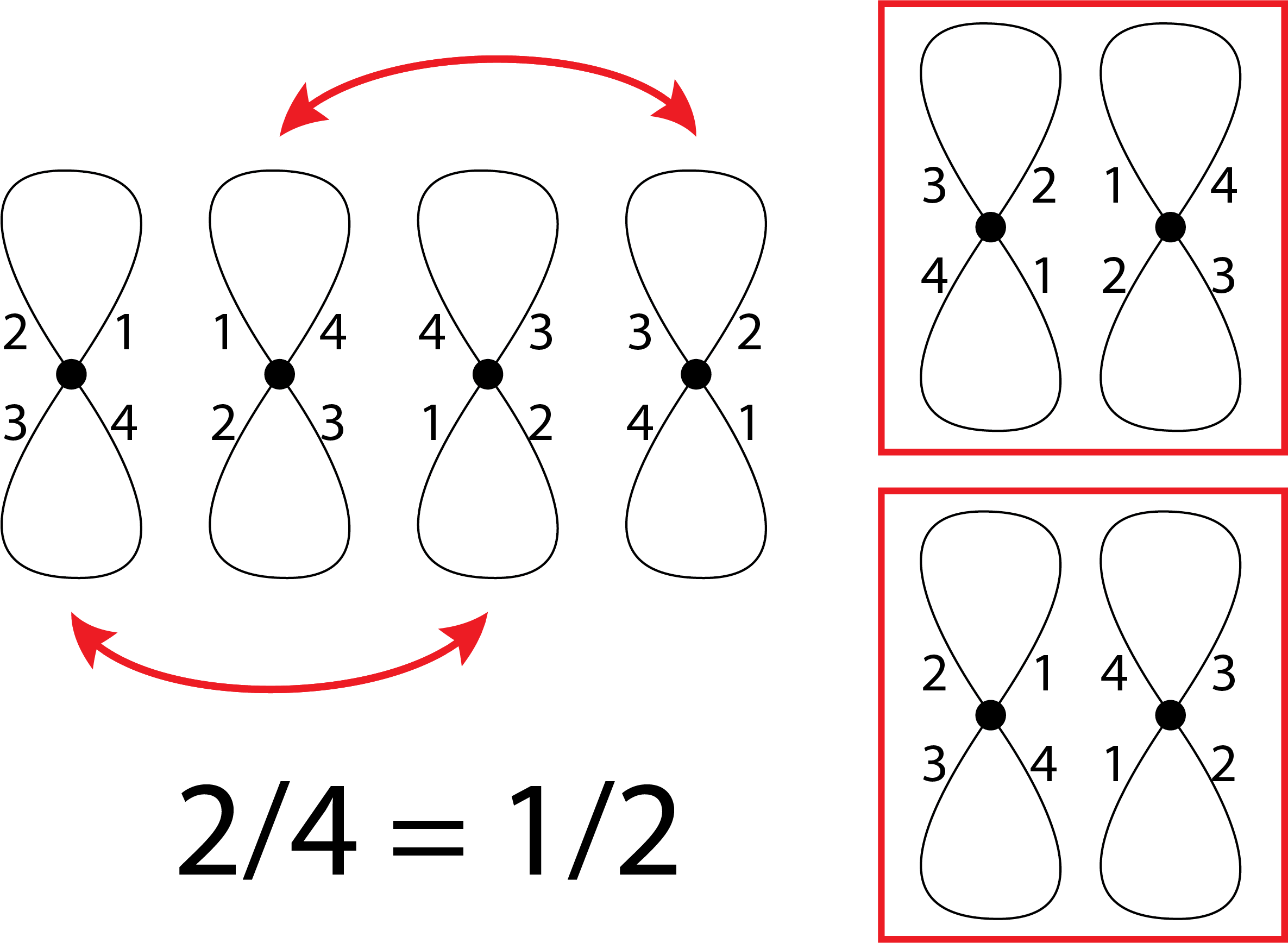}
\caption{Labeled maps with 1 vertex on a surface of genus 0.}
\label{Fig:map_counts}
\end{figure}

Given a compact oriented surface of genus $g$, enumerating maps means finding how many \emph{different} maps with $j$ vertices can be embedded in the surface. To address the issue of permuting vertices, a common approach is to label them ahead of time. Accounting for all of the vertex labelings therefore introduces a factor of $j !$ in the map count. To describe the graph, one also needs to label the \emph{darts}, which are the half-edges emanating from each vertex. Pairing two darts defines an edge between the vertices to which they are attached. Because the surface is oriented and the vertices are already numbered, the order in which the dart labels are arranged about each vertex is fixed. In the case of a regular-valence map, such that the degree of each vertex is $2 \nu$, labeling the darts therefore results in an additional factor of $(2 \nu)^j$, where the $2 \nu$ comes from rotating the dart numbers counterclockwise about each vertex. When enumerating maps, one then needs to specify whether counts are for \emph{labeled maps} or \emph{unlabeled maps}, the former being $(2 \nu)^j \cdot j!$ as large as the latter. Another consideration that affects map counts is related to symmetries of the surface. Indeed, two maps are considered equivalent (i.e. not different) if there is an orientation-preserving
homeomorphism of the surface onto itself that leaves the map
unchanged. This may lead to unlabeled counts that are \emph{rational} numbers. An illustrative example is provided in Figure \ref{Fig:map_counts}. On a surface of genus 0, there are \emph{a priori} four 4-valent labeled maps with one vertex. These are shown on the left-hand side of Figure \ref{Fig:map_counts} and are obtained from one-another by rotating counterclockwise the labeling of the 4 darts. However, because the plane (or sphere) is invariant under rotations, the first and third maps  are the same. So are the second and fourth (red double arrows). There are therefore two \emph{different} labeled maps, one for each box on the right-hand-side of the figure. Consequently, the number of different unlabeled maps is $2 / (4^1 \cdot 1!) = 1/2$, which is not integral.

In some cases, topological frustration will lead to map counts that are zero. As for instance explained in \cite{bib:elt22}, the Euler characteristic $\chi$ of the cellular polyhedron determined by a $2\nu$-valent map with $j$ vertices on a surface of
genus $g$ is given by
\[
\chi = 2 - 2g = j - E + F = j - \nu j + F \ge j - \nu j + 1,
\]
where $E = \nu j$ is the number of edges (or half of the number of darts) of the map and $F \ge 1$ is the number of faces of the polyhedron obtained by cutting the surface along the graph.
This implies that $j \ge (2g-1) / (\nu-1)$, which means that on a surface of genus $g$, there are no maps with a number of vertices strictly less than $(2g-1)/ (\nu-1)$.

In parallel with the above-mentioned developments on the topic of map enumeration, the study of \emph{random matrix theory} as a zero dimensional gauge field theory led to  an equivalent problem, through application of Feynman diagrammatic expansions. Bessis, Itzykson, and Zuber \cite{bib:biz80} recognized the connection and were able to bring ideas from statistical mechanics to bear on the question of map enumeration. The centerpiece of this approach is the random matrix partition function for Hermitian random matrix ensembles. Specifically, the Gibbs measure for this ensemble is of the form 
\begin{equation} \label{Gibbs}
d\mu_{\bf t} = \frac{1}{Z_N^{(n)}(\bf t)} \exp \left(- N\, \text{Tr}(V_{{\bf t}}(M))\right) dM,
\end{equation}
where
\begin{equation}
V_{{\bf t}}(\lambda) = \left(\frac1{2} \lambda^2 + \sum_{k=1}^J t_k \lambda^k\right),
\end{equation}
and $V_{{\bf t}}$ has a vector of $J$ parameters ${\bf t}=(t_1, \cdots, t_J) \in \mathbb{R}^J$ with $J$ even.
The normalization
\begin{eqnarray} \label{partition}
  Z_N^{(n)}(\bf t) &=&  
  \int_{\mathcal{H}_n}  
  \exp \left(- N \, \text{Tr}(V_{{\bf t}}(M))\right) dM,
\end{eqnarray}        
referred to as the \emph{partition function} of the ensemble, makes $d\mu_{\bf t}$ a probability measure on the space $\mathcal{H}_n$ of $n \times n$ Hermitian matrices. 
The statistical mechanical problem is to determine the asymptotic behavior of this partition function in the limit as the matrix size $n$ tends to infinity, along with the parameter $N$, so that $n/N$ remains fixed at a value near 1. Such a limit is analogous to more traditional thermodynamic limits in which the  area of a lattice tends to infinity, an idea that goes back to 't Hooft \cite{bib:tH74} in the general gauge theory setting. The ratio $n/N$ is referred to as the \emph{'t Hooft parameter} and is normally denoted by $x$ in the literature. However, to avoid any confusion with the notation in the rest of this manuscript, we will call it $\alpha$. 

There are at least two approaches for formally assessing this limit. One is based on a direct iterative study of the 
random matrix resolvent (Green's function) associated to the measure
\eqref{Gibbs} and was first explored by Ambjorn, Chekov, Kristjansen, and Makeenko \cite{bib:ackm93}. This is usually referred to as the \emph{loop equation} approach. The other is based on a clever idea due to Dyson \cite{bib:dy62} and was first exploited in this setting by \cite{bib:biz80}.  The key is to note that due to its unitary conjugation invariance, the partition function may be reduced to a multiple integral over eigenvalues,

\[
Z_N^{(n)}(\mathbf{t}) = {\mathcal K}_N^{(n)}\, \widehat{Z}_N^{(n)}(\mathbf{t}),
\]
where
\begin{equation} \label{eq:eigen}
  \widehat{Z}_N^{(n)}(\mathbf{t}) =
  \bigints \cdots \bigints \exp 
  \left\{ -N^2 \left[ \frac{1}{N} \sum_{j=1}^n V_{{\bf t}}(\lambda_j) - \frac{1}{N^2} \sum_{j \ne \ell} \log |\lambda_j-\lambda_\ell| \right] \right\}  d^n \lambda,
\end{equation} 
and the multiplicative factor ${\mathcal K}_N^{(n)}$, related to the eigenvectors, is independent of $\mathbf{t}$. 
The two versions of the partition function, \eqref{partition} and \eqref{eq:eigen} therefore differ only by an overall constant. Since we will only be considering the ratios
\[
Z_N^{(n)}({\bf t})/Z_N^{(n)}(0) = \widehat{Z}_N^{(n)}({\bf t})/\widehat{Z}_N^{(n)}(0),
\]
the two versions are, for our purposes, effectively equivalent. 

An influential early study of map enumeration that used methods  of random matrix theory concerned the enumeration of one-face maps (or, dually, maps with just one vertex). It emerged tangentially from calculations of the Euler characteristics of moduli spaces of Riemann surfaces \cite{bib:hz86}. With this one-face constraint, the generating function was seen to be polynomial in $1/n^2$. A remarkable conjecture that emerged from this and some of the other studies around that time (see \cite{bib:biz80} and references therein) was that in the large $n$ limit, ${\mathcal G}_N^{(n)} = \log \big(Z_N^{(n)}({\mathbf t})/Z_N^{(n)}(0) \big)$ has, in some sense, an asymptotic expansion of the form
\begin{equation}
    {\mathcal G}_N^{(n)} =  n^2 e_0({\mathbf t},\alpha) + e_1({\mathbf t},\alpha) + 
    \frac{1}{n^2} e_2({\mathbf t},\alpha) + \cdots + \frac{1}{n^{2g-2}} e_g({\mathbf t},\alpha) + \cdots
    \label{eq:e_g_exp}
\end{equation}
in inverse powers of $n^2$, whose coefficients $e_g({\mathbf t},\alpha)$ are, when evaluated at $\alpha=1$, generating functions for the enumeration of labeled $g$-maps (i.e. maps that can be embedded in a surface of genus $g$). This conjecture was prompted by the observation that, when $n$ is finite and $\alpha$ is near 1, the coefficients of the MacLaurin expansion of ${\mathcal G}_N^{(n)}$ in $\mathbf t$ are, at each order, polynomial in $1/n^2$. Rearranging terms in the formal series obtained as $n \to \infty$ led to expression \eqref{eq:e_g_exp}.
However, this has potential issues. In particular, the resulting expansion, in which successive coefficients of $1/n^{2g-2}$ are obtained by re-summing an infinite number of terms in $\mathbf t$, may not be asymptotically valid. Another issue concerns contributions that are beyond all orders in the $n^{-2}$-gauge. While such terms do not prima facie affect the asymptotic expansion, which is not unique, they do play a role in getting the exact representation of the large $n$ limit, as we shall see shortly. A rigorous derivation of \eqref{eq:e_g_exp} therefore requires evaluating ${\mathcal G}_N^{(n)}$ in the limit as $n \to \infty$ with $\alpha$ near 1, and proving that the coefficients remain, in that limit, generating functions. In other words, one needs to demonstrate that the $e_g(\alpha,{\mathbf t})$ appearing in the asymptotic expansion of ${\mathcal G}_N^{(n)}$ as $n \to \infty$ are infinitely differentiable at $\mathbf{t} = 0$ when $\alpha \simeq 1$, and that their MacLaurin series agree with the expressions obtained for finite $n$, thereby preserving the significance of each coefficient in terms of map counts. Conjecture \eqref{eq:e_g_exp} was proved by Ercolani and McLaughlin in \cite{bib:em03}, by rigorously establishing an even broader class of asymptotic limits. This was done by using the alternative representation of ${\mathcal G}_N^{(n)}$ based on \eqref{eq:eigen} and making essential use of orthogonal polynomials, as described below.

Dyson had observed that by replacing the van der Monde determinant (written as the exponential of the sum of the log terms) appearing in \eqref{eq:eigen} by a determinant of \emph{orthogonal polynomials} associated to the exponential weight 
\begin{equation} \label{eq:weight}
w(\lambda) = e^{- N V_{\bf t}(\lambda)}, \quad 
V_{\bf t}(\lambda) = \left(\frac1{2} \lambda^2 + \sum_{k=1}^J t_k \lambda^k\right),
\end{equation}
the iterated integral in \eqref{eq:eigen} reduced to an $n$-fold product of one-dimensional integrals, where $n$ is the matrix size. The authors of \cite{bib:em03} applied Riemann-Hilbert analysis to get an exact representation of ${\mathcal G}_N^{(n)}$ as the integral of an appropriate test function against an explicit density of the form
\begin{eqnarray} \label{eq:density}
K_{{\bf t},n}(\lambda, \eta)= e^{ -(1/2) ( V_{{\bf t}, N}(\lambda) + V_{{\bf t},N}(\eta)   ) } \sum_{\ell = 0}^{n-1} \pi_\ell(\lambda) \pi_\ell(\eta),
\end{eqnarray} 
where the
monic orthogonal polynomials $\big\{\pi_n\big\}$ are defined on the real line with respect to the exponential weight $w(\lambda)$ of Equation \eqref{eq:weight}. The sum in \eqref{eq:density} is oscillatory with the density of the oscillations increasing with $n$. Nevertheless the limit can be understood in the sense of weak convergence of measures and analyzed in detail. Although integrals of test functions against the limiting measure are \emph{far from} manifestly having a $1/n^2$ expansion, an order by order analysis shows that, remarkably, this is the case. Establishing such a result requires, roughly speaking, a careful asymptotic separation of the dominant oscillatory behavior from exponentially decaying pieces that will lead to terms beyond all orders. Doing this was one of the fundamental results of \cite{bib:em03}.

Later, integrating other test functions of polynomial type against the density \eqref{eq:density} led Ercolani, McLaughlin, and Pierce \cite{bib:emp08} to establish other asymptotic expansions relating orthogonal polynomials to enumerative generating functions. In particular, they proved that as $n \to \infty$, the asymptotic expansion
\begin{equation}
\label{eq:gen_exp}
b_n^2 = \alpha \left(z_0(t,\alpha) + \frac{1}{n^2} z_1(t,\alpha) + \cdots + \frac{1}{n^{2 g}} z_g(t,\alpha) + {\mathcal O}\left(\frac{1}{n^{2 (g+1)}}\right)\right),
\end{equation}
for $\alpha$ near 1 is uniformly valid on compact sets in $t \in \mathbb{C}$, where $t$ has positive real part. Here, the ${\mathbf t}$ of \eqref{eq:weight} is one-dimensional and equal to $t \equiv t_J$; all of the vertices therefore have the same even valence, $J = 2 \nu$. The coefficients $b_n^2$ appearing in \eqref{eq:gen_exp} relate the
monic orthogonal polynomials $\big\{\pi_n\big\}$ through the recurrence relation 
\[
\lambda\, \pi_n(\lambda) = \pi_{n+1}(\lambda) + b_n^2\, \pi_{n-1}(\lambda).
\]
On the right-hand-side of \eqref{eq:gen_exp}, the $z_g(t, \alpha)$ are generating functions for counts of 2-legged maps on surfaces of genus $g$, where \emph{2-legged} means that in addition to the $j$ $J$-valent vertices, the map has two vertices of degree one, called legs. In Section \ref{sec:OPPCM}, we explain how a dynamical systems analysis of the discrete evolution equation for the coefficients $x_n = b_n^2$ leads to an asymptotic expansion that can be compared with \eqref{eq:gen_exp} to obtain explicit expressions for the $z_g(t,\alpha)$ in terms of $z_0(t,\alpha)$. The existence of such expressions, as well as their rational form, stems from the work of Ercolani \cite{bib:er11,bib:er14}, which is summarized below.

In \cite{bib:er11}, Ercolani showed that each generating function $z_g(t,\alpha)$ is a \emph{rational function} of $z_0(t,\alpha)$,
\begin{equation}
\label{eq:rat_form_z}
z_g(t,\alpha) = \frac{z_0 (z_0 - 1) P_{3 g - 2}(z_0)}{(\nu - (\nu - 1) z_0)^{5 g - 1}},
\end{equation}
where $z_0 = z_0(t,\alpha)$ and $P_{3g-2}$ is a polynomial of degree $3 g - 2$. This is \textit{a priori} surprising since $z_g$, which may be used to count maps with an arbitrary number of vertices, only depends on $z_0$ (which is explicitly known from the string equation \eqref{eq:string_eq} below) and a \emph{finite} number of parameters: the half-valence $\nu$ and the $3g - 1$ coefficients of $P_{3g-2}$. The connection with the generating functions $e_g$ was made in \cite{bib:er14}, which established that $e_g(t,\alpha)$ solves a Cauchy-Euler equation in $t$, with a forcing term that depends on the generating functions $\{z_k(t,\alpha), \ k \le g\}$ and $\{e_k(t,\alpha), \ k < g\}$. Building on this result, \cite{bib:er14} demonstrated that for $g \ge 2$, $e_g(t,\alpha)$ is also a rational function of $z_0 = z_0(t,\alpha)$,
\begin{equation}
\label{eq:rat_form}
e_g(z_0)= \dfrac{(z_0-1)^r
Q_{5g-5-r}(z_0)}{\left(\nu - (\nu-1) z_0\right)^{5g-5}}, \qquad r = \max\left\{1, \left\lfloor \dfrac{2 g - 1}{\nu - 1} \right\rfloor \right\},
\end{equation}
where $Q_{5g-5-r}$ is a polynomial of degree $5g-5-r$. This, which proved and extended a conjecture proposed for $\nu = 2$ in \cite{bib:biz80}, led to the partial fraction expansion
\begin{equation}
e_g(z_0)=C^{(g)}+\sum_{\ell=0}^{3g-3}\dfrac{q_\ell^{(0)}}{(\nu-(\nu-1) z_0)^{2g+\ell-2}}, \quad C^{(g)}, q_\ell^{(0)} \in \mathbb{R}.
\label{eq:eg}
\end{equation}
In this context, the number of labeled $2\nu$-valent maps with $j$ vertices that can be embedded in a surface of genus $g$ is given by
\[
{\mathcal N}_{2\nu,e}(g,j) = \left. (-1)^j \, \dfrac{d^j e_g}{d t^j}\right\vert_{t=0},
\]
where, $e_g$ is a function of $z_0$ and $z_0$ solves what is known as the \emph{(continuum) string equation}
\begin{equation}
\label{eq:string_eq}
1 = z_0 + c_\nu t\, z_0^\nu, \qquad c_\nu = 2 \nu \binom{2 \nu - 1}{\nu - 1}.
\end{equation}

These results made it possible to express the number of regular even-valent $g$-maps with $j$ vertices in terms of $z_0$ and a finite number of real coefficients (specifically $3 g + j - 3$). Indeed, taking successive derivatives of \eqref{eq:eg} with respect to $t$ gives, for $j \ge 1$, \cite{bib:er14}
\begin{equation}
\label{eq:parfrac_exp}
e_g^{(j)}(z_0) := (-1)^j\, \dfrac{d^j e_g}{d t^j} = c_\nu^j\, z_0^{j \nu + 1} \sum_{\ell = 0}^{3 g + j - 4} \dfrac{q_\ell^{(j)}}{\left(\nu - (\nu-1) z_0\right)^{2g+\ell+j-1}}.
\end{equation}
Recently, we showed \cite{bib:elt23b} that such a formulation provides an explicit expression for the counts in terms of hypergeometric functions. Specifically, because the $e_g$ are analytic in a neighborhood of $t = 0$, one may write
\[
{\mathcal N}_{2\nu,e}(g,j) =
\dfrac{j!\, c_\nu^j}{2\pi i}\oint \dfrac{e_g(\eta)}{\eta^{j+1}} d\eta,
\]
where $\eta = - c_\nu t$. An evaluation of the above integral leads to
\begin{align}
\label{eq:eg_hyper}
{\mathcal N}_{2\nu,e}(g,j) = j!\, c_\nu^j (\nu - 1)^j  \sum_{\ell = 0}^{3g-3} & \left(q_\ell^{(0)}
\binom{(2g-4) + (\ell+j)}{j}\right.  \\
& \left. \ {_2}F_1
\left( \genfrac{}{}{0pt}{}
{-j,\ 1- \nu j}{4-2g-(\ell+j)}; \frac{1}{1-\nu} \right)\right), \nonumber
\end{align}
for $g \ge 2$, where ${_2}F_1$ denotes the \emph{Gauss hypergeometric function}. Although the right-hand-side of \eqref{eq:eg_hyper} is a closed-form formulation involving the genus $g$, the number of vertices $j$, and the valence $2 \nu$ of each vertex, an explicit expression that does not rely on hypergeometric functions may be obtained in the special case when $\nu = 2$, as described in Sections \ref{sec:vector_recurrence} and \ref{sec:4-valent}.

Assuming an exact closed expression for the counts for general $j$ is known, it is natural to ask about the asymptotic behavior of these counts as $j \to \infty$. When no closed form is available, this can sometimes be read off from the recursive structure of counts; as mentioned earlier such an approach goes back to Bender et al. The rational expressions for generating functions found in \cite{bib:er11,bib:er14} enable one to determine this asymptotic behavior for {\em all} values of $g$ even when the specific coefficients in those expressions are not known. We refer to \cite{bib:elt23b} for details on this.

The discussion so far addressed the enumeration of regular, even-valent maps. It is however natural to ask about the case when $J$ is odd. In this situation, the weight of Equation \eqref{eq:weight} has a divergent integral on the real line and, consequently, the previously discussed methods involving orthogonal polynomials are no longer meaningful. Important progress was made by Bleher and Dea\~no
\cite{bib:bd13} for the case when $J=3$, by using non-Hermitian orthogonal polynomials. These are defined
by deforming the contour of integration for \eqref{eq:weight} from the real line to an appropriate contour in the complex plane along which the total integral converges. Although such a transformation leads to complex expressions, asymptotic representations such as \eqref{eq:gen_exp} remain valid and the analysis can be carried out as before. Combinatorially, setting $J = 3$ corresponds to enumerating maps that are surface triangulations.
In \cite{bib:ep12}, Ercolani and Pierce went on to explicitly calculate the analogue of \eqref{eq:eg}  
for low values of the genus. 
Extending this type of analysis to the case of regular odd valence $J$ has additional complications; how to handle these, and obtain formulations for asymptotic map counts, is described by Ercolani and Waters in \cite{bib:ew22}.

Other, more purely combinatorial recursive methods for studying map enumeration have been developed since the publication of \cite{bib:em03} in 2003. In particular we refer to \cite{bib:cms09}, \cite{bib:bg12}, and the references therein. We note as well a fairly recent paper by Dubrovin and Yang \cite{bib:dy17}, which extends the result of Harer and Zagier \cite{bib:hz86} 
from one-face maps to maps with a fixed finite number of faces (or, dually, vertices).  This introduces a recursion based on a \emph{Lax operator} that also arose earlier in connection with the orthogonal polynomial resolvent operator introduced by Ercolani and Waters \cite{bib:wa15,bib:ew22}.

The remainder of this article is organized as follows. Section \ref{sec:dyn_perspectives} describes how adding a dynamical perspective to the results of \cite{bib:em03,bib:emp08,bib:er11,bib:er14} leads to explicit formulations for the generating functions $z_g$ and $e_g$ in terms of $z_0$ \cite{bib:tip20,bib:elt22,bib:elt23a}, as well as to non-recursive expressions for four-valent map counts \cite{bib:elt23b}. Section \ref{sec:new_methodology} explains how all of these advances may be combined into a general framework for map enumeration \cite{bib:tip20,bib:elt23a}. Section \ref{sec:mixed_case} presents new partial results on the mixed valence case where both $t_3$ and $t_4$ are non-zero, corresponding to tessellations with triangles and quadrilaterals. In conclusion, Section \ref{sec:conclusions} discusses open challenges and future directions.

\section{Dynamical perspectives} \label{sec:dyn_perspectives}
As summarized in the introduction of this article, the topic of map enumeration, which started as a combinatorial question, greatly benefited from concepts associated with the fields of statistical physics, random matrices, and orthogonal polynomials. Recently, we discovered that bringing a dynamical lens to some of the remaining open problems in this area could foster further advances. This section explains how ideas from dynamical systems theory, both linear and nonlinear, lead to explicit expressions for the generating functions $z_g(t,\alpha)$ and $e_g(t,\alpha)$ and, in the 4-valent case, to explicit formulas for map counts. We focus on the broad ideas and refer the interested reader to \cite{bib:elt22,bib:elt23a,bib:elt23b} for proofs and further details. 

\subsection{Orthogonal polynomials, Painlev\'e equation, and center manifold} \label{sec:OPPCM}

As introduced in Section \ref{sec:intro}, the genus expansion \eqref{eq:gen_exp} relates the generating functions $z_g(t,\alpha)$ and the recurrence coefficients associated with orthogonal polynomials defined on the real line in terms of the weight \eqref{eq:weight}, with $\mathbf t$ real. Remarkably, these coefficients satisfy evolution equations of the Painlev\'e type. For instance, \cite{bib:elt22} considers the family of orthonormal polynomials $\{p_i\ \vert\ \text{deg}(p_i)=i\}$ such that
\[
\int_{\mathbb{R}} p_n(\lambda) p_m(\lambda) w(\lambda) d \lambda = \delta_{nm}, \quad w(\lambda) = \exp\left(-N \left( \dfrac12 \lambda^2 + \dfrac{r}4 \lambda^4\right)\right).
\]
Orthogonality implies that the polynomials $p_i$ are related to one another via an equation of the form
\begin{equation}\label{eq:threetermrecquartic}
\lambda\, p_n(\lambda) = b_{n+1}\, p_{n+1}(\lambda) + b_n\, p_{n-1}(\lambda),    
\end{equation}
where the coefficients $b_n$ are the same as in Equation \eqref{eq:gen_exp} and solve the difference equation
\[
r\,b_n^2\left( b_{n+1}^2 + b_n^2 + b_{n-1}^2 \right) + b_n^2 = \dfrac{n}{N}.
\]
Setting $x_n = b_n^2$ leads to the following form of the \emph{discrete Painlev\'e I Equation} (dP1) \cite{bib:va18,bib:cm20}, also referred to as the (discrete) string equation
\begin{equation}
\label{eq:dP1} 
x_{n+1}+x_n+x_{n-1} = \dfrac{n}{N\,r\,x_n}-\dfrac{1}{r}, \quad n \in \mathbb{N},\ x_n \in \mathbb{R}.
\end{equation}

Our first  dynamical perspective consists in viewing \eqref{eq:dP1} as a discrete dynamical system and studying it as such \cite{bib:elt22}. Indeed, dP1 may be written as the 3-dimensional, discrete, nonlinear, autonomous, first-order dynamical system
\begin{align} \label{eq:dP1_dyn_sys} 
\begin{split}
x_{n+1} &= \dfrac{n_n}{N\,r\,x_n} -\dfrac{1}{r}-x_n-y_n \\ 
y_{n+1} &= x_n \\
n_{n+1} &= n_n + 1 \quad (\text{with } n_n = n).
\end{split}
\end{align}
The coefficients $b_n^2$ follow a particular orbit of this dynamical system. This trajectory is defined by the Freud recurrence \cite{bib:fre76} and has the following initial condition, which in turn depends on the first and second moments of the exponential weight,
\[
x_1 = \dfrac{\mu_2}{\mu_0}, \quad y_1 = 0, \quad
\mu_i=\int_{-\infty}^\infty \lambda^i \exp\left(-N\left(\lambda^2/2+(r/4)\lambda^4\right)\right) d\lambda.
\]
For $N = 1$ and $r = 1$, a high precision numerical estimation of $x_1$ leads to \cite{bib:elt22}
\[
x_1 = \dfrac{\mu_2}{\mu_0} \simeq 0.47.
\]
In addition, we know that the Freud orbit diverges as $n \to \infty$ according to \cite{bib:fre76}
\begin{equation} \label{eq:freudasymp}
    x_n \sim \sqrt{\dfrac{n}{3Nr}} \ \text{as } n \to \infty.
\end{equation}   

It is therefore natural to introduce a transformation that brings this ``fixed point at infinity'' to finite space. To this end, in \cite{bib:elt22}, we define the change of coordinates $(x,y,n) \to (s, f, u)$, where
\[
s = \frac{y}{x} + 1 + \frac{1}{r x}, \quad f = \frac{n}{N r x^2} - \frac{y}{x}, \quad u = - \frac{1}{r x}.
\]
Since $x_n \to \infty$ as $n \to \infty$ for the Freud orbit, the corresponding sequence $\{u_n\}$ converges to $0$. In addition since $x_n\sim y_n$ the sequence $\{s_n\}$
converges to $2$, and using \eqref{eq:freudasymp} we see that the sequence $\{f_n\}$ converges to $2$ as well. In other words, the above transformation has the property that the Freud orbit converges to the fixed point $P_\infty$ with coordinates $(2,2,0)$ in the $(s,f,u)$ space. 

An analysis of \eqref{eq:dP1} written in $(s,f,u)$ coordinates, which reads 
\begin{align*}
s_{n+1} &= Z_n f_n, \qquad Z_n = (u_n + f_n -1)^{-1}\\
f_{n+1} &= Z_n^2 \left(s_n + \gamma u_n^2 \right)\\
u_{n+1} &= Z_n u_n,
\end{align*}
reveals \cite{bib:elt22} that $u = 0$ is an \emph{invariant plane} that contains the only two (real) fixed points of this dynamical system, namely the origin and $P_\infty$. Linearization about $P_\infty$ indicates that the three eigenvalues are $\lambda = 1$ and $\lambda_\pm = -(2 \pm \sqrt 3)$: one direction is neutral, one is contracting, and one is expanding. Orbits that converge to $P_\infty$ generically converge along the slowest direction, in this case the center direction. If the Freud orbit converged to $P_\infty$ along its exponentially contracting direction, the exponential dependence in $n$ would preclude the existence of an asymptotic expansion in powers of $n^{-1/2}$, which we know exists \cite{bib:mnz85,bib:bmn88,bib:ansva15}. Consequently, the Freud orbit must converge along the center direction, a statement which is confirmed by the numerical evidence presented in \cite{bib:elt22}. This simple fact provides a method to find the asymptotic expansion of the Freud orbit in powers of $n^{-1/2}$ as $n \to \infty$. Indeed, the center manifold theorem \cite{bib:i79} allows us derive a Taylor expansion for the center manifold, which is a curve $\mathcal C$ that is invariant under dP1 and tangent to the center direction of the linearization at $P_\infty$. We find \cite{bib:elt22}
\begin{align} \label{eq:s_f_infty}
\begin{split}
s_\infty(u)&=2-u-\frac{\gamma}{6} u^2-\frac{\gamma}{36} u^3-\frac{\gamma  (3 \gamma +1)}{216} u^4+{\mathcal O}(u^{5}) \\
f_\infty(u)&=2-u+\frac{\gamma}{6} u^2+\frac{\gamma}{36} u^3-\frac{\gamma  (3 \gamma -1)}{216} u^4+{\mathcal O}(u^{5})
\end{split}
\end{align}
as $u \to 0.$ This expansion, which is asymptotic, can be \emph{continued to arbitrary order} in powers of $u$. Going back to the initial variables, 
\[
x=-\frac{1}{r u}, \quad y = - \frac{s+u-1}{r u}, \quad \frac{n}{N} = \frac{s+f+u-1}{r u^2},
\]
the following implicit expression for $u_n$ in terms of $n$ is obtained along the center manifold:
\[
\frac{n}{N} = \frac{s_\infty(u_n)+f_\infty(u_n)+u_n-1}{r u_n^2}.
\]
This in turn leads to an asymptotic expansion for $x_n$ along the Freud orbit, of the form \cite{bib:elt23a}
\[
b_n^2 = x_n \sim \sum_{k=-1}^\infty \dfrac{c_k}{n^{k/2}},
\]
so that 
\[
\left| b_n^2 -\sum_{k=-1}^m \dfrac{c_k}{n^{k/2}} \right| < \dfrac{K_m(N,r)}{n^{(m+1)/2}}, \ \ \  m\geq-1.
\]
Here, the constant $K_m$ depends on the truncation order $m$ of the Laurent expansion of $b_n^2$ and on the parameters $N$ and $r$ that appear in the exponential weight $w$. We call this asymptotic expansion the \emph{center manifold expansion}. 

 An analysis of dP1 as a discrete, autonomous dynamical system therefore allows us to derive an asymptotic expansion of the Freud orbit by calculating the Taylor expansion of the center manifold associated with $P_\infty$, which is the non-trivial fixed point of \eqref{eq:dP1} written in $(s,f,u)$ coordinates. Separately, we know that the Freud orbit corresponds to the sequence of coefficients $b_n^2$, which are related to the generating functions $z_g(t,\alpha)$ through the genus expansion \eqref{eq:gen_exp}. Finding a region of common asymptotic validity for the center manifold and genus expansions therefore provides a systematic method to obtain explicit expressions for the $z_g(t,\alpha)$ in terms of $z_0$. We implemented this approach in \cite{bib:elt23a}, which yielded expressions for $z_g(z_0)$ and subsequently $e_g(z_0)$ in the 4-valent case, with $1 \le g \le 7$.

Converting an equation of Painlev\'e type, here dP1, into a first-order autonomous system with finite fixed points and using the center manifold theorem to obtain an asymptotic expansion of the Freud orbit sets up a very general framework to study the $1/n$ dependence of the coefficients $b_n^2$. This methodology can naturally be extended to other systems of orthogonal polynomials, as long as the weight $w(\lambda)$ has the form given in \eqref{eq:weight} and the fixed point at infinity has \emph{only one} center direction. An example is provided in Section \ref{sec:mixed_case_cm}, in the case of maps of mixed (3 and 4) valence. A drawback of course is that the dimension of the dynamical system increases with the degree of, or the number of non-zero coefficients $t_i$ appearing in, $w(\lambda)$.

\subsection{Vector recurrence} \label{sec:vector_recurrence}
We now return to the expression for the number of labeled $2 \nu$-valent $g$-maps with $j$ vertices,
\[
{\mathcal N}_{2\nu,e}(g,j) = \left. (-1)^j \, \dfrac{d^j e_g}{d t^j}\right\vert_{t=0}.
\]
The main advantage of the rational form expression provided in \eqref{eq:rat_form} is that the successive derivatives of $e_g$ can easily be calculated once the coefficients $q_\ell^{(j)}$ appearing in the partial fraction expansion shown in \eqref{eq:parfrac_exp} are known. Indeed, since $z_0 = 1$ when $t = 0$, we have for $j \ge 1$
\begin{equation}
\label{eq:map_counts_q_l}
{\mathcal N}_{2\nu,e}(g,j) = \left. (-1)^j\, \dfrac{d^j e_g}{d t^j} \right \vert_{t = 0} = c_\nu^j\, \sum_{\ell = 0}^{3 g + j - 4} q_\ell^{(j)}.
\end{equation}
Moreover, implicit differentiation of the string equation
\[
1 = z_0 + c_\nu \, t \, z_0^\nu, \qquad c_\nu = 2 \nu \binom{2 \nu - 1}{\nu - 1},
\]
with respect to $t$ shows that $dz_0 / dt$ is a function of $z_0$ only:
\[
\dfrac{d z_0}{d t} = - c_\nu \, \dfrac{z_0^{\nu+1}}{\nu - (\nu - 1) z_0}.
\]
Consequently, combining the above expression with the $t$-differentiation of \eqref{eq:parfrac_exp} leads to a two-step recurrence equation for the coefficients $q_\ell^{(j)}$ \cite{bib:er14,bib:elt23b},
\[
q_\ell^{(j)}=\beta_{j,\ell} \ q_{\ell-1}^{(j-1)} - \alpha_{j,\ell} \ q_\ell^{(j-1)}, \qquad \quad 0 \le \ell \le 3 g - 4 + j, \quad j \ge 2,
\]
where $\beta_{j,\ell} = \nu \big(2 g+\ell+j-3 \big)$, $\alpha_{j,\ell} = \big(2 g + \ell - 2 -(\nu - 1)(j-1)\big)$ and $q_{-1}^{(j)} = 0$. For $j \ge 1$, we can thus define a semi-infinite vector of coefficients
\[
V^{(j)}=c_\nu^j \left[q_0^{(j)}, q_1^{(j)}, \cdots, q_\ell^{(j)}, \cdots,  q_{3g-4+j}^{(j)}, 0, \cdots\right]^T,
\]
whose dynamics is given by
\[
V^{(j)} = c_\nu^{j-1} \left(\overleftarrow{\prod_{n=2}^j} M^{(n)}\right) \ V^{(1)}, \qquad \overleftarrow{\prod_{n=2}^j} M^{(n)} = M^{(j)} \, M^{(j-1)} \cdots M^{(2)},
\]
where the bidiagonal matrices $M^{(j)}$ depend on the coefficients $\alpha_{j,\ell}$ and $\beta_{j,\ell}$. In addition, the vector $V^{(j)}$ has only a finite number of non-zero entries, consistent with the expression given in \eqref{eq:map_counts_q_l} \cite{bib:elt23b}.

This dynamic representation is advantageous because it reduces finding the map counts to summing a finite number of non-zero entries of a semi-infinite vector, obtained by successive applications of a non-autonomous (i.e. $j$-dependent) linear map. The resulting orbit, defined as the sequence of points $\{V^{(j)}\}$ in the associated semi-infinite space, is parametrized by the genus $g$, and leads to \emph{recursive} knowledge of the counts $\{{\mathcal N}_{2\nu,e}(g,j)\}$ as soon as the initial condition $V^{(1)}$ is known. The question of finding an explicit expression for the solution of this linear, non-autonomous, infinite-dimensional dynamical system in terms of its initial condition is currently open, except in the 4-valent case, where $\nu = 2$. Indeed, a remarkable simplification occurs in this situation, which allows us to reduce the dynamics of $\{V^{(j)}\}$ to that of a finite dimensional non-autonomous linear dynamical system, whose solutions can be calculated \cite{bib:elt23b}. This approach is summarized in the next section. Regarding how to find the initial condition $V^{(1)}$, the work of \cite{bib:elt23a} explains how to obtain $e_g$ as a function of $z_0$, and therefore the initial vector of coefficients $V^{(1)}$, by equating rescaled versions of the center manifold and genus expansions in a region of common asymptotic validity.

\subsection{The 4-valent case} \label{sec:4-valent}
Although the number, $3 g + j - 3$, of possibly non-zero entries of $V^{(j)}$ increases with $j$, some of these coefficients may nevertheless vanish. As shown in \cite{bib:elt23b}, when $\nu = 2$, the dynamics of the sequence $\{V^{(j)}\}$ simplifies: for $g \ge 2$, all of the potentially non-zero entries of $V^{(j)}$ are consecutively found in a band whose length, $5 g - 5$, is independent of $j$. Consequently, the dynamics of the sequence $\{V^{(j)}\}$ can be recast into that of a \emph{finite-dimensional} discrete dynamical system by writing
\[
V^{(j)} = \left[0,0, \dots, x_1^{(j)}, x_2^{(j)}, \dots, x_{5 g - 5}^{(j)}, 0, \dots \right]^T
\]
where the $x_k^{(j)}$ represent the entries of $V^{(j)}$ that are possibly non-zero, and defining the $(5g-5)$-dimensional map
\[
V_{e,g}^{(j+1)} = c_2 A^{(j)} V_{e,g}^{(j)},
\]
where $c_2 = 12$ and $V_{e,g}^{(j)}$ is obtained from $V^{(j)}$ by setting
\[
V_{e,g}^{(j)} = \left[x_1^{(j)}, x_2^{(j)}, \dots, x_{5 g - 5}^{(j)}\right]^T.
\]
An expression for the matrix $A^{(j)}$ is provided in  Appendix C.2 of \cite{bib:elt23b}. Since counts of maps with $j$ vertices are obtained from summing the entries of $V^{(j)}$, we have
\[
{\mathcal N}_{2\nu,e}(g,j) = \left. (-1)^j\, \dfrac{d^j e_g}{d t^j} \right \vert_{t = 0} = c_\nu^j\, \sum_{\ell = 0}^{3 g + j - 4} q_\ell^{(j)} = \sum_{i=1}^\infty V^{(j)}[i] = \sum_{i=1}^{5 g - 5} V_{e,g}^{(j)}[i],
\]
where the bracket $[i]$ refers to the $i^\text{th}$ entry of the corresponding vector.

Whereas the semi-infinite matrices $M^{(n)}$ and $M^{(j)}$ do not commute when $n \ne j$, it turns out that the matrices $A^{(n)}$ and $A^{(j)}$ always do \cite{bib:elt23b}! Writing 
\[
V_{e,g}^{(j)} = c_2^{j-1} \left(\prod_{k=1}^{j-1}
A^{(k)} \right) V_{e,g}^{(1)},
\]
and defining the row vector ${\mathcal R}^{(j)}$ as the column sums of the matrix
\[
{\mathcal M}^{(j)} = \prod_{k=1}^{j-1}
A^{(k)},
\]
we can write the map counts as the contraction
\[
{\mathcal N}_{4,e}(g,j) = \sum_{i=1}^{5 g - 5} V_{e,g}^{(j)}[i] = c_2^{j-1}\ {\mathcal R}^{(j)} \cdot V_{e,g}^{(1)}.
\]
Since the matrices $A^{(k)}$ commute, they share a basis of eigenvectors, which makes it possible to calculate their product ${\mathcal M}^{(j)}$ and consequently the entries of the row vector ${\mathcal R}^{(j)}$. As detailed in \cite{bib:elt23b},
\[
{\mathcal R}^{(j)}[n] = \frac{1}{2^{n-1}} \sum_{k=1}^{n} \left[\binom{n-1}{k-1} \prod_{\ell=1}^{j-1}
2 (2 \ell + k)\right].
\] 
Knowledge of the initial vector of coefficients $V_{e,g}^{(1)}$ is therefore sufficient to obtain exact expressions for the 4-valent counts ${\mathcal N}_{4,e}(g,j)$, for all values of $j$, when $g \ge 2$. As explained earlier, the method developed in \cite{bib:elt23a} provides expressions for $e_g(z_0)$. Taking one derivative with respect to $t$ and expanding the result in powers of $1/(2-z_0)$ leads to the initial vector of coefficients $V^{(1)}$ and subsequently to $V_{e,g}^{(1)}$. This last step requires padding the vector of non-zero coefficients obtained from $V^{(1)}$ with a prescribed number of zeros on the left. Details, as well as the initial vectors $V_{e,g}^{(1)}$ for $2 \le g \le 7$  are provided in \cite{bib:elt23b} (appendix F). As an example of counts obtained in \cite{bib:elt23b} with this method, we have
\begin{align*}
{\mathcal N}_{4,e}(5,j) = & 12^{j -1} \left(\prod_{k=1}^7(j-k)\right) \\ &
\Bigg(4^{j -1} j ! \left(\frac{38213}{1146617856} j^{2}+\frac{915313}{2293235712} j -\frac{1940327}{53508833280}\right)\\
  & \ \ -\frac{\left(2 j \right)!}{j !} \left(\frac{211033}{2319969600} j^{2}+\frac{8139013}{71455063680} j +\frac{1}{887040}\right)\Bigg)
 \end{align*}
 for all $j \ge 1$. Few map count formulas, found in \cite{bib:biz80} for $g=2$ and \cite{bib:bgm21} for $g=3$, were known prior to the work of \cite{bib:elt23b}, which provides \emph{non-recursive}, explicit expressions  for up to $g = 7$, for both regular and 2-legged maps.

\section{A new methodology for map enumeration} \label{sec:new_methodology}

\begin{figure}[ht]
\includegraphics[width=.9 \linewidth]{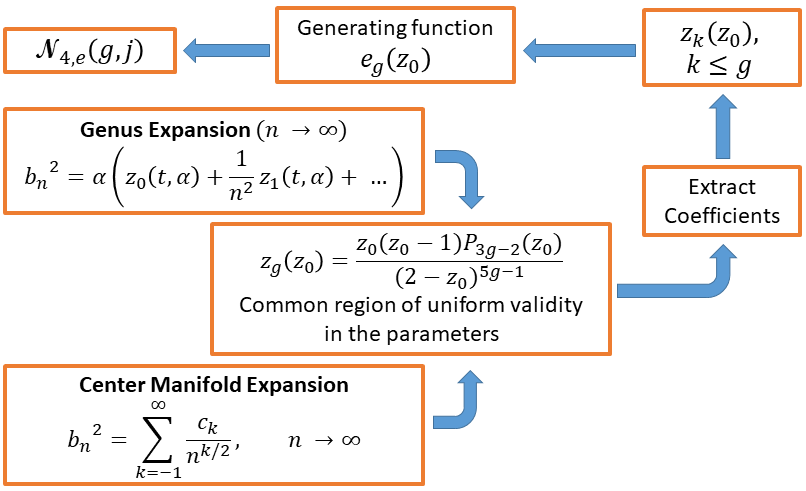}
\caption{The methodology proposed in \cite{bib:elt22,bib:elt23a,bib:elt23b} to obtain counts of 4-valent maps.
}
\label{Fig:flow_chart}
\end{figure}

The advances of \cite{bib:elt22,bib:elt23a,bib:elt23b} summarized in the previous section build on the results of \cite{bib:em03,bib:emp08,bib:er11,bib:er14} to establish a general methodology for map enumeration. There are four key ingredients central to this approach. The first one is the existence of a \emph{genus expansion} \cite{bib:emp08} that identifies the generating functions for 2-legged maps as the coefficients of an asymptotic expansion of the $b_n^2$ in powers of $1/n^2$ for large $n$. The second ingredient is an independent, explicit formulation of how the $b_n^2$ behave as $n \to \infty$; this is provided by the \emph{center manifold expansion} \cite{bib:elt22}. Third, the existence of a rational form \cite{bib:er11} giving $z_g(z_0)$ in terms of $z_0$ guarantees that only a \emph{finite} number of coefficients need to be identified to explicitly write each $z_g$ as a function of $z_0$. Finally, there needs to be a \emph{common region of asymptotic validity} of the (suitably transformed) genus and center manifold expansions. Comparing the two leads to a set of equations for the coefficients appearing in the rational expression of $z_g$ in terms of $z_0$, whose unique solution provides an explicit formulation of $z_g(z_0)$ \cite{bib:elt23a}. Expressions for the generating functions $e_g(z_0)$ may then be uniquely determined by solving an appropriate Cauchy-Euler equation \cite{bib:er14,bib:elt23a}. Once the generating functions are known, the map counts can be calculated through differentiation and evaluation at $t = 0$. In some cases, non-recursive, explicit expressions in terms of the genus $g$ and the number of vertices $j$ may be obtained \cite{bib:elt23b}. This framework is illustrated in the flow chart of Figure \ref{Fig:flow_chart}, in the case of 4-valent maps. The four pillars supporting this methodology, namely the genus expansion, the center manifold expansion, the rational form expression for $z_g$, and the ability to extract the coefficients of the rational form by comparing the two expansions, appear in the bottom part of the figure. These elements combine to provide explicit expressions for $z_g(z_0)$ (rightmost box on the first row), which in turn lead to knowledge of $e_g(z_0)$ and of the map counts ${\mathcal N}_{4,e}(g,j)$.

The program we just described has so far only been fully implemented for 4-valent maps \cite{bib:tip20,bib:elt22,bib:elt23a,bib:elt23b}. However, many partial results suggest it can be generalized to regular- and mixed-valence map counts. For instance, although a center manifold expansion has not been established in the case of even valence other than 4, the results of \cite{bib:mnz85} prove the existence of an asymptotic expansion of the Freud orbit as $n \to \infty$ for general even valence ${\mathcal V} = 2 \nu$. Similarly, the existence of the genus expansion \eqref{eq:gen_exp} and of the rational form \eqref{eq:rat_form_z} are known in that case. In addition, all of the elements needed to compare the center manifold and genus expansions were proven in \cite{bib:elt23a} for ${\mathcal V} = 2 \nu$. 

Also of interest is the mixed-valence case, corresponding to a vector $\mathbf t$ in Equation \eqref{eq:weight} with more than one dimension. Section \ref{sec:mixed_case} presents the dynamical systems perspective associated with mixed 3- and 4-valent maps, and shows that the corresponding center manifold expansion is a regular perturbation of the 4-valent center manifold expansion obtained in \cite{bib:elt22}, as long as $t_4 \ne 0$.

\section{Preliminary results on the mixed 3- and 4-valent case}
\label{sec:mixed_case}
In this section, we derive the discrete dynamical system for the recurrence coefficients of the orthonormal polynomials associated with the potential 
\begin{equation}
\label{eq:mixed_pot}
V_{\bf t}(\lambda) = \frac1{2} \lambda^2 + t_3 \lambda^3 + t_4 \lambda^4, \quad t_3 \ne 0, \ t_4 > 0.
\end{equation}
We then show that, after an appropriate change of coordinates, the resulting autonomous dynamical system has a fixed point $P_\infty$ with only one center direction, and provide numerical evidence that the Freud orbit converges to that fixed point. Finally, we obtain an expression for the associated center manifold which, for small values of $t_3$, is a regular perturbation of Equations \eqref{eq:s_f_infty}.

\subsection{Derivation of the mixed valence discrete dynamical systems} 
\label{sec:mixed_case_deriv}

Consider the class of orthonormal polynomials $\{ p_i \}$ defined by
\[
\int_{\mathbb{R}} p_n(\lambda) p_m(\lambda) w(\lambda) d \lambda = \delta_{nm}, \quad w(\lambda) = \exp\left(-N V_{\bf t}(\lambda) \right),
\]
where $V_{\bf t}(\lambda)$ is given in Equation \eqref{eq:mixed_pot}. This section provides a short summary of known results for this exponential weight. We refer the reader to \cite{bib:mag86} for more detailed derivations of the general statements.

With $V_{\bf t}$ containing odd degree terms, the even symmetry witnessed in \eqref{eq:threetermrecquartic} is lost and the three-term recurrence for the polynomials takes the form
\[
\lambda\, p_n(\lambda) = b_{n+1}\, p_{n+1}(\lambda) +a_{n}\, p_{n}(\lambda) + b_n\, p_{n-1}(\lambda).
\]
This relationship can be compactly expressed as $\lambda  \mathbf{p} = {\mathcal J} \mathbf{p}$, where
\[\mathbf{p} =\begin{bmatrix}
p_0\\
p_1  \\
 p_2\\
\vdots
\end{bmatrix}, 
\quad
{\mathcal J} = \begin{bmatrix}
a_0 & b_1 &  &  \\
b_1 & a_1 & b_2 &  \\
 & b_2 & a_2 & b_3 \\
 &  &  & \ddots
\end{bmatrix},  \]
and the semi-infinite matrix ${\mathcal J}$ is called the \emph{Jacobi matrix}.  Combining orthogonality conditions with integration by parts leads to the following relations \cite{bib:mag86}: 
\begin{align}\begin{split}    
\label{eq:genfre}
  b_n P'({\mathcal J})_{n,n-1} &= n/N \\
  P'({\mathcal J})_{n,n} &= 0.   
 \end{split}
\end{align}
In the above, given a matrix $A$, we have used $A_{n,m}$ to denote the $n,m$-th entry of $A$ \emph{with rows and columns indexed starting at 0}.

For $V_{\bf t}$ given in Equation \eqref{eq:mixed_pot}, system \eqref{eq:genfre} takes the form
\begin{align} \label{eq:genfre_mixed}
\begin{split}
  n/N = b_n^2 & + 4t_4b_n^2(a_{n-1}^2+a_n^2+a_{n-1}a_n +b_n^2+b_{n-1}^2+b_{n+1}^2)\\ 
  & + 3t_3b_n^2(a_{n-1}+a_n) \\
 0 = a_n & + 4t_4(a_n^3+a_{n-1}b_n^2+a_{n+1}b_{n+1}^2+2a_nb_n^2+2a_nb_{n+1}^2)\\
 & + 3t_3(a_n^2+b_n^2+b_{n+1}^2),
\end{split}\end{align}
which is a second order non-autonomous discrete dynamical system for the variables $x_n = b_n^2$ and $a_n$. 

\subsection{Center manifold}
\label{sec:mixed_case_cm}
Setting
\begin{align*}
&t_{3} = \frac{\sqrt{\zeta}}{3},\ t_{4} = \frac{1}{4 \eta},\ a_{n} = \sqrt{\zeta}z_{n},  \\
&b_{n}^{2} = x_{n},\ x_{n-1} = y_{n},\ z_{n -1} = w_{n},
\end{align*}
in \eqref{eq:genfre_mixed} leads to the first-order, 4-dimensional, non-autonomous dynamical system
\begin{align*}
\begin{split}
& x_{n +1} = 
\left(\left(-w_{n}-z_{n}\right) \eta -w_{n}^{2}-w_{n} z_{n}-z_{n}^{2}\right) \zeta +\left(-1+\frac{n}{N x_{n}}\right) \eta -x_{n}-y_{n}, \\
& y_{n +1} = x_{n}, \\
& z_{n +1} = 
\left(-\frac{\eta  z_{n}^{2}}{x_{n +1}}-\frac{z_{n}^{3}}{x_{n +1}}\right) \zeta +\left(-\frac{x_{n}}{x_{n +1}}-1-\frac{z_{n}}{x_{n +1}}\right) \eta -\frac{x_{n} w_{n}}{x_{n +1}}-\frac{2 x_{n} z_{n}}{x_{n +1}}-2 z_{n}, \\
& w_{n +1} = z_{n}.
\end{split}
\end{align*}
When $t_3$ is small, the above system is a regular perturbation of the dP1 equation introduced in Section \ref{sec:OPPCM}, for which $a_n = 0$. This can be seen by setting $\zeta = 0$ in the above equations. Although $z_{n} = 0$ is not a solution in that case, the relation $a_n = \sqrt \zeta z_{n}$ does imply $a_n = 0$. The first-order autonomous version of dP1, given in Equations \eqref{eq:dP1_dyn_sys}, is therefore recovered (with $\eta = 1/r$) in the limit as $t_3 \to 0$. In contrast, setting $t_4 = 0$ is a singular limit since this corresponds to sending $\eta$ to infinity.

\begin{figure}[ht]
\includegraphics[width=.95\linewidth]{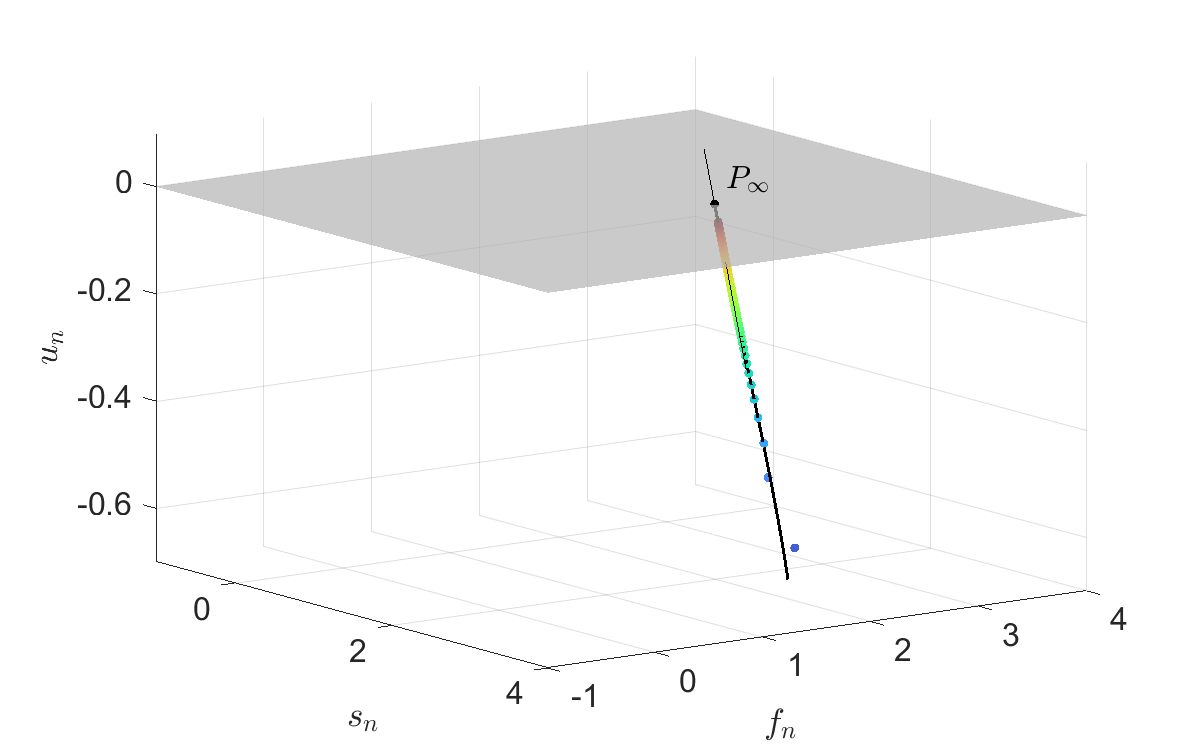}\\
\includegraphics[width=.95\linewidth]{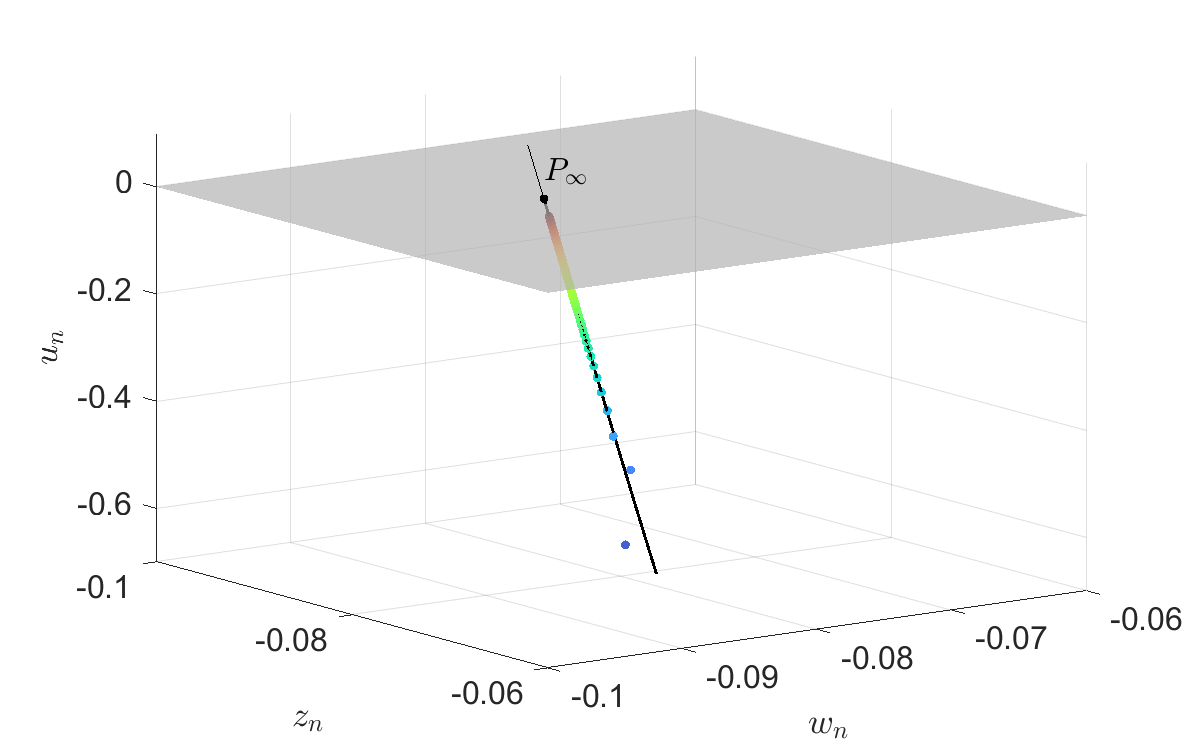}
\caption{Numerical simulation of the Freud orbit (dots) for the mixed valence case and plot of the center manifold of $P_\infty$ (solid curve) expanded to fourth order in $u$. The dynamics is shown in $(s,f,u,z,w)$ coordinates. The dots are colored according to the value of the iteration index $n \in \{2, 3, \cdots, 700\}$, with blue (resp. red) corresponding to low (resp. high) indices. The Freud orbit converges to $P_\infty$, which is on the invariant hyperplane $u=0$ in grey).}
\label{Fig:cm}
\end{figure}

As before, we introduce a change of coordinates that brings the ``fixed point at infinity'' to finite space,
\[
s = \frac{y}{x}+1+\frac{\eta}{x}, \quad f = 
\frac{n \eta}{N \,x^{2}}-\frac{y}{x}, \quad u = -\frac{\eta}{x},
\]
which in turns leads to the following 5-dimensional autonomous dynamical system
\begin{align}
s_{n +1} & = \frac{\zeta Q_{n}+f_{n} Z_{n}}{\zeta Q_{n}+1}, \notag\\ 
f_{n +1} & = 
-\frac{Z_{n} \left(\left(-\gamma  u_{n}^{2}-s_{n}\right) Z_{n}+\zeta Q_{n}\right)}{\left(\zeta Q_{n}+1\right)^{2}}
,\notag\\ 
u_{n +1} & = \frac{u_{n} Z_{n}}{\zeta Q_{n}+1}, \label{eq:mixed_DS}\\
z_{n +1} & = \frac{1}{\eta  \left(\zeta Q_{n}+1\right)}
\Big(\left(\eta  Z_{n} u_{n} z_{n}^{2}+Z_{n} u_{n} z_{n}^{3}-\eta^{2} Q_{n}-2 \eta  Q_{n} z_{n}\right) \zeta \notag \\
& \qquad \qquad \qquad \quad +\left(\left(u_{n}-2\right) Z_{n}-2\right) \eta  z_{n}+\left(-Z_{n}-1\right) \eta^{2}-\eta  Z_{n} w_{n}\Big),\notag\\
w_{n +1} & = z_{n}, \notag
\end{align}
where $\gamma = 1/(\eta N)$ and the quantities $Z_n$, $Q_n$, and $H_n$ are defined by
\[
Z_{n} = \left(u_{n}+f_{n}-1\right)^{-1}, \ 
Q_{n} = \frac{1}{\eta} \left(Z_{n} H_{n} u_{n}\right), \ 
H_{n} = 
\eta  w_{n}+\eta  z_{n}+w_{n}^{2}+w_{n} z_{n}+z_{n}^{2}.
\]
An analysis of System \eqref{eq:mixed_DS} with Maple \cite{bib:maple} shows that it has one real fixed point $P_\infty$ with $(s,f,u,z,w)$ coordinates
\[
P_\infty = \left(2,2,0,-\frac{\eta}{3},-\frac{\eta}{3}\right),
\]
and a line of real fixed points through the origin, given by 
\[
s = 0, \quad f = 0, \quad u = 0, \quad w = z.
\]
Linearization about $P_\infty$ reveals that the five eigenvalues are $\lambda = 1$ with multiplicity one, and $\lambda = -2 \pm \sqrt 3$, each with multiplicity two. Consequently, $P_\infty$ has a one-dimensional center direction. A center manifold expansion may be obtained by writing $s,\ f,\ z$, and $w$ in powers of $u$, and solving order by order after demanding that the resulting curve be invariant under the dynamics of system \eqref{eq:mixed_DS} and tangent to the center direction at $P_\infty$. Such a calculation can easily be performed with the assistance of a computer algebra software. Using Maple \cite{bib:maple}, we obtain the following asymptotic expansion as $u \to 0$,
\begin{align*}
\begin{split}
s_\infty(u) & = 2-u-\frac{1}{6} \gamma  u^{2}+\left(\frac{1}{108} \eta  \gamma  \zeta -\frac{1}{36} \gamma \right) u^{3}-\frac{1}{1944} \gamma  \left(27 \gamma +\left(\eta  \zeta -3\right)^{2}\right) u^{4}\\
& \quad +\mathrm{O}\! \left(u^{5}\right)
, \\
f_\infty(u) & = 2+\left(\frac{\eta  \zeta}{3}-1\right) u+\frac{1}{6} \gamma  u^{2}+\left(-\frac{\eta^{3} \zeta^{3}}{2187}+\frac{\eta^{2} \zeta^{2}}{243}-\frac{\eta  \left(\gamma +1\right) \zeta}{108}+\frac{\gamma}{36}\right) u^{3}\\
& \quad +\left(\frac{\eta^{4} \zeta^{4}}{19683}-\frac{4 \eta^{3} \zeta^{3}}{6561}+\frac{\left(\gamma +\frac{14}{3}\right) \eta^{2} \zeta^{2}}{1944}-\frac{\eta  \left(\gamma +1\right) \zeta}{324}-\frac{\gamma^{2}}{72}+\frac{\gamma}{216}\right) u^{4}\\
& \quad +\mathrm{O}\! \left(u^{5}\right),
\end{split}
\end{align*}
together with
\begin{align*}
\begin{split}
 z_\infty(u) & = -\frac{\eta}{3}+\left(\frac{1}{81} \eta^{2} \zeta -\frac{1}{18} \eta \right) u+\left(-\frac{1}{1458} \eta^{3} \zeta^{2}+\frac{5}{972} \eta^{2} \zeta -\frac{1}{108} \eta \right) u^{2}\\
 & \quad +\frac{1}{52488} \eta  \left(2 \eta  \zeta -9\right) \left(\eta^{2} \zeta^{2}-6 \eta  \zeta -27 \gamma +9\right) u^{3}\\
 & \quad -\frac{23 \left(\eta  \zeta -\frac{9}{2}\right)}{12754584} \left(\eta^{3} \zeta^{3}-9 \eta^{2} \zeta^{2}-\frac{2187}{23} \zeta \left(\gamma -\frac{8}{27}\right) \eta+\frac{6561 \gamma}{23}-\frac{729}{23}\right) \eta  u^{4}\\
 & \quad +\mathrm{O}\! \left(u^{5}\right), \\ 
w_\infty(u) & = -\frac{\eta}{3}+\left(\frac{1}{81} \eta^{2} \zeta -\frac{1}{18} \eta \right) u+\left(-\frac{1}{1458} \eta^{3} \zeta^{2}+\frac{5}{972} \eta^{2} \zeta -\frac{1}{108} \eta \right) u^{2}\\
& \quad +\frac{1}{52488} \eta \left(2 \eta  \zeta -9\right) \left(\eta^{2} \zeta^{2}-6 \eta  \zeta +27 \gamma +9\right) u^{3}\\
& \quad -\frac{23 \left(\eta  \zeta -\frac{9}{2}\right)}{12754584} \left(\eta^{3} \zeta^{3}-9 \eta^{2} \zeta^{2}+\frac{2187}{23} \zeta \left(\gamma +\frac{8}{27}\right) \eta -\frac{6561 \gamma}{23}-\frac{729}{23}\right) \eta u^{4}\\
& \quad +\mathrm{O}\! \left(u^{5}\right).
\end{split}
\end{align*}
Figure \ref{Fig:cm} shows the Freud orbit for $t_3=t_4=1$. Its initial condition
\[
x_2 \simeq 0.3933, \quad y_2 \simeq 0.3210, \quad z_2 \simeq -0.0791, \quad w_2 \simeq -0.0803.
\]
was obtained with Mathematica 12 \cite{bib:mathematica} after setting $t_3=t_4=1$ in the defining measure \eqref{eq:mixed_pot}. The first few moments of this measure were numerically approximated using 800 digits of precision and the initial conditions are simple rational functions in these moments. Successive iterates were then found through application of the recurrence \eqref{eq:genfre_mixed} at that same level of precision. As illustrated in Figure \ref{Fig:cm}, the Freud orbit (dots) converges to $P_\infty$ along the center manifold approximation (solid line)
\[
s=s_\infty(u),\quad f=f_\infty(u),\quad z=z_\infty(u), \quad w=w_\infty(u)
\]
parametrized by $u$. As was the case for dP1 \cite{bib:elt22}, the agreement is very good, even for low values of $n$. Using the relation
\[
n = \eta N \frac{f_\infty(u_n) + s_\infty(u_n) + u_n - 1}{u_n^2}
\]
leads to an expression for $u_n$ in terms of $n$ along the center manifold, which can then be used to find asymptotic expansions for $x_n$ and $a_n$ on the Freud orbit as $n \to \infty$.

To complete the program of Figure \ref{Fig:flow_chart} and derive map counts for the (3,4) mixed valence case, a proof of the existence of the rational form $z_g(z_0)$, and an exact understanding of its structure, are needed. In addition, the numerical evidence presented here needs to be supplemented by a dominant balance argument showing that, in $(s,f,u,z,w)$ coordinates, the Freud orbit converges to $P_\infty$. Our preliminary investigations indicate that these tasks, as well as a comparison of the genus and center manifold expansions, can indeed be performed. This however goes beyond the scope of the present article and is left for future work.

\section{Conclusions: open problems and challenges}
\label{sec:conclusions}
Map enumeration is a fascinating area of research that stands at the intersection of statistical physics, random matrices, orthogonal polynomials, and discrete dynamical systems theory. This article highlights a framework, developed over the years by the first author and his collaborators, that leads to the rigorous derivation of generating functions for map counts. The corresponding methodology, which relies on four pillars, is described in Section \ref{sec:new_methodology} and illustrated in Figure \ref{Fig:flow_chart}. The dynamical systems perspective summarized in Section \ref{sec:dyn_perspectives} addresses one of these pillars and has allowed the present authors to systematically derive recursive and non-recursive expressions for the number of $g$-maps of regular even valence. Here, \emph{recursive} means that explicit expressions for the generating functions $e_g(z_0)$ are formulated and that map counts are evaluated by taking successive derivatives as the number of vertices $j$ of the map increases. In contrast, non-recursive map counts are expressions that explicitly refer to the parameter $j$. Table \ref{tab:pillars} summarizes the progress made so far in the context of the framework of Figure \ref{Fig:flow_chart}, including the new contribution of Section \ref{sec:mixed_case}. Table \ref{tab:results} shows the associated results, as well as how they compare to previous works available in the literature.

\begin{table}[ht]
\begin{center}
\begin{tabular}{|m{2.1 cm} || m{2.1 cm} | m{2.1 cm} | m{2.1 cm}| m{2.1 cm} |} 
 \hline
 \hfil Valence \hfil & Genus \newline Expansion & Center \newline Manifold \newline Expansion & Rational \newline Form & Common \newline Validity of 2 \newline Expansions \\ 
 \hline
 \hfil 4 \hfil & \checkmark \hfil \cite{bib:emp08} & \checkmark \hfil \cite{bib:elt22} & \hfil \checkmark \hfil \newline \cite{bib:er11,bib:er14} & \hfil \checkmark \hfil \cite{bib:elt23a} \\ 
 \hline
 \hfil $2 \nu,\ \nu \in \mathbb{N}$ \hfil & \checkmark \hfil \cite{bib:emp08} & Existence of \newline expansion is \newline known \newline \cite{bib:mnz85} & \hfil \checkmark \hfil \newline \cite{bib:er11,bib:er14} & \checkmark \hfil \cite{bib:elt23a} \\
 \hline
 \hfil 3 \hfil & Existence \newline established \cite{bib:bd13} \hfil & & \hfil In progress \hfil & \ \newline \\
 \hline
 \rule{0pt}{15pt} Mixed 3 \& 4 & \rule{0pt}{15pt} \hfil In progress \hfil & \rule{0pt}{15pt} \hfil \checkmark \hfil {\bf [\S \ref{sec:mixed_case_cm}]} \hfil & \rule{0pt}{15pt} \hfil In progress \hfil & \rule{0pt}{15pt} \hfil In progress \hfil \\[10pt]
 \hline
 \end{tabular}
 \end{center}
  \caption{The four pillars (Columns 2-5) of the methodology described in Section \ref{sec:new_methodology} and Figure \ref{Fig:flow_chart} for different valences (Column 1). A check mark indicates a successful derivation of the corresponding pillar for the given valence. \label{tab:pillars}}
\end{table}

\begin{table}[hbtp]
\begin{center}
 \begin{tabular}{|m{1.9 cm} || m{2 cm} | m{2 cm} | m{2 cm}| m{2 cm} |} 
 \hline
 \hfil Valence \hfil & \hfil 4 \hfil & \hfil $2 \nu,\ \nu \in \mathbb{N}$ \hfil & \hfil 3 \hfil &  \hfil Mixed 3 \& 4 \hfil \\ 
 \hline\hline
 Results Related to $\, $ Map Counts & $\bullet\ 0 \le g \le 2$ \cite{bib:emp08}
 $\bullet\ 0 \le g \le 7$ (Recursive) \cite{bib:elt23a}
$\bullet\ 1 \le g \le 7$ (Explicit) \cite{bib:elt23b} & 
$\bullet\ 0 \le g \le 2$ \cite{bib:emp08} $\ $
$\bullet\ 2 \le g \quad$ (Hyper- $\quad$ geometric functions) \cite{bib:elt23b} & $\bullet\ 0 \le g \le 2 \ $ \cite{bib:ep12} & \hfil In progress \hfil \\ 
 \hline
 \hline
 Previously Known Results Related to $\ $ Map Counts & $\bullet\ 0 \le g \le 2$ \cite{bib:biz80} 
 $\bullet\ 0 \le g \le 5$ \cite{bib:vp} $\qquad$
 $\bullet\ 0 \le g \le 5$ \cite{bib:dy17} $\ \ $
 $\bullet\ g=3 \qquad$ \cite{bib:bgm21} &
 &
 $\bullet\ 0 \le g \le 2$ \cite{bib:bd13}
 $\bullet\ 0 \le g \le 5$ \cite{bib:dy17}
 & \\
 \hline
 \end{tabular}
 \end{center}
  \caption{Map counts associated with the four pillars of the methodology described in Section \ref{sec:new_methodology} and Figure \ref{Fig:flow_chart} (Row 2), and results obtained from other approaches (Row 3), for different valences (Row 1). \label{tab:results}}
\end{table}

There are many other directions that can be pursued in the context of the present program. Of immediate interest are triangulations (3-valent maps) and the mixed valence case corresponding to tessellations of surfaces with triangles and quadrangles, mentioned in Section \ref{sec:mixed_case}. Indeed, the 3-valent case is a singular limit of the mixed (3,4)-valent problem, whereas the limit towards the well-studied 4-valent case is, as discussed above, regular. In addition, during our exploration of map enumeration, we came across striking numerical evidence that suggests remarkable conjectures, some of which are relevant to other areas of mathematics. We briefly discuss each of these topics below. 

\subsection{3-valent case}
It is natural to inquire if the methodology we described in Section \ref{sec:new_methodology} 
will extend to the regular 3-valent setting. We fully expect that will be the case. One already knows from \cite{bib:bd13} that a genus expansion in the regular 3-valent setting exists. There are two pillars that remain to be completed so that all elements are available. 

\subsubsection{Rational forms}
The first missing pillar is the derivation of rational expressions for the $e_g$. For low genera, this was done in \cite{bib:ep12}, resulting in, for example, 
\begin{align*}
e_{0}\! \left(t \right) & = 
\frac{\ln \! \left(z_0 \! \left(t \right)\right)}{2}+\frac{\left(z_0 \! \left(t \right)-1\right) \left(z_0 \! \left(t \right)^{2}-6 z_0 \! \left(t \right)-3\right)}{12 z_0 \! \left(t \right)+12}, \\
e_{1}\! \left(t \right) & = 
-\frac{1}{24} \ln \! \left(\frac{3}{2}-\frac{z_0 \left(t \right)^{2}}{2}\right), \\
e_{2}\! \left(t \right) & = 
\frac{\left(z_0 \! \left(t \right)^{2}-1\right)^{3} \left(4 z_0 \! \left(t \right)^{4}-93 z_0 \! \left(t \right)^{2}-261\right)}{960 \left(z_0 \! \left(t \right)^{2}-3\right)^{5}},
\end{align*}
where $z_0(t)$ is implicitly defined by the string equation
\[
1 = z_0 \! \left(t \right)^{2}-72\, t^{2} z_0 \! \left(t \right)^{3}.
\]
The extension of this to rational forms of general genus is in progress.

\subsubsection{Center manifold expansion}
The second required pillar to complete is the center manifold expansion. To derive the associated dynamical system, consider the class of orthonormal polynomials $\{ p_i \}$ defined by
\[
\int_{\Gamma} p_n(\lambda) p_m(\lambda) w(\lambda) d \lambda = \delta_{nm}, \quad w(\lambda) = \exp\left(-N P(\lambda)\right),
\]
where  $P(\lambda)=\dfrac12 \lambda^2 + t_3 \lambda^3$  and $\Gamma$ is the union of the positive real axis with the ray $\{re^{4 \pi i /5}\}_{ r>0}$, chosen so that these integrals converge.
In this particular case, the recurrence equations for the $a_n$ and $b_n$ coefficients take the form:
\begin{align}\begin{split} \label{eq:cubicfreab}
  b_n^2 + 3t_3b_n^2(a_{n-1}+a_n)  &= n/N \\
 a_n + 3t_3(a_n^2+b_n^2+b_{n+1}^2)  &= 0. 
\end{split}\end{align}
For $n$ fixed, Equations \eqref{eq:cubicfreab} define an autonomous \emph{Quispel-Roberts-Thompson mapping}, or QRT mapping for short. The dynamics of this class of integrable systems is defined geometrically in terms of an iteration involving the intersection of lines with a biquadratic. Each of these is an initial-condition-dependent invariant curve. Together, they foliate the phase plane. The QRT parameters are encapsulated in the two matrices:
\begin{equation}
    A_0 = \begin{bmatrix}
        0 & 0 & -t_3/3 \\
         -t_3/3 & -1 & 0 \\
          0 & (n+1)/N &r \\
    \end{bmatrix} 
\quad
        A_1 = \begin{bmatrix}
        0 & 0 & 0 \\
         0 & 0 & 0 \\
          0 & 0 &1 \\
    \end{bmatrix}
\end{equation}
As a non-autonomous dynamical system, \eqref{eq:cubicfreab} is Painlev\'e \cite{bib:sak01} and exhibits singularity confinement. For further details on how the QRT mapping informs the Painlev\'e dynamics, see \cite{bib:tip20}.

Let $x_n$ denote $b_n^2$, as before. Then Equations \eqref{eq:cubicfreab}  lead to the following first order recurrence for $a_n$ and $x_n$:
\begin{align}\begin{split}\label{eq:cubicfre}
  x_{n+1} &= -\frac{a_n}{3t_3}-a_n^2-x_n \\
 a_{n+1} &=\frac{n+1}{3Nt_3 x_{n+1}}-\frac{1}{3t_3}-a_n.
\end{split}\end{align}
Knowledge of the leading order asymptotic behavior of $x_n$ and $a_n$ along the Freud orbit will suggest a suitable change of coordinates that will make the dynamics converge to a fixed point. If the linear stability analysis of that fixed point reveals the existence of a one-dimensional center direction, then a systematic derivation of the center manifold expansion will be possible. Work along these lines, which includes establishing a proper definition of the Freud orbit, is currently in progress.

\subsection{Conjectures and open questions}

As mentioned above, while carrying out the program of Section \ref{sec:new_methodology} for maps of regular even valence, we identified a collection of facts, both numerical and theoretical, which we summarize below. In addition to providing interesting problems for curious mathematicians, these highlight the range of potential applications of the present work.

\bigskip
\subsubsection{An explicit identity involving the Gauss hypergeometric function} A comparison of the expression for map counts of even valence in terms of hypergeometric functions (Equation \eqref{eq:eg_hyper} above) with their explicit expression in the case when $\nu = 2$ led us to the following combinatorial conjecture.
\begin{conj} \label{conj:combinatorics}
\cite{bib:elt23b} For integer values of $g \ge 1$, $\ell \ge 0$, and $j \ge 1$, the following identity is true
\begin{eqnarray*}
&& j!\, 2^{\ell + 2 g - 1} \binom{(2g-2) + (\ell+j)}{j}
\ {_2}F_1 \left( \genfrac{}{}{0pt}{}
{-j,\ -2 j}{2-2g-(\ell+j)}; -1 \right)\\
&=& \sum_{k=1}^{\ell+2 g} \binom{\ell + 2 g - 1}{k-1} \prod_{m=0}^{j-1} 2 (2 m + k).
\end{eqnarray*}
\end{conj}
\noindent We have numerically verified the validity of Conjecture \ref{conj:combinatorics} for values of $\ell$ between 0 and 20, and values of $g$ and $j$ between 1 and 20. This conjecture was recently proven by M.L. Yattselev \cite{Ya24}.

\bigskip
\subsubsection{Singularity confinement and period-3 orbits of dP1}
We previously indicated that, when written in $(s,f,u)$ coordinates, dP1 admits $u=0$ as an invariant plane. Restricting the dynamics of dP1 to the $u=0$ plane gives a two-dimensional discrete dynamical system, which reads
\[
s_{n+1} = Z_n f_n, \quad f_{n+1} = - s_n Z_n^2, \qquad Z_n = (f_n - 1)^{-1}.
\]
In \cite{bib:elt22}, we pointed out that the dynamics of polar orbits of dP1 appear to shadow  this system as soon as $|x_n|$ is large enough. In addition, we suggested an association between singularity confinement in dP1 and the line $s+f = 1$ of period-3 orbits (which includes one singular orbit) of the above two-dimensional dynamical system. Elucidating these connections, in particular the relation between the distance of non-polar orbits from the plane $u=0$ and singularity confinement, remains an open challenge.

\bigskip
\subsubsection{Interlacing zeros of generating functions for 2-legged maps} When $\nu = 2$, the rational form expression \eqref{eq:rat_form_z} of $z_g$ in terms of $z_0$ simplifies into
\[
z_g=\dfrac{z_0(z_0-1)^{2 g} S_{g-1}(z_0)}{(2-z_0)^{5g-1}},
\]
where $S_{g-1}$ is a polynomial of degree $g-1$. In \cite{bib:elt23a}, we established such expressions for values of $g$ up to 7 and noticed that the zeros of the polynomials $S_{g-1}$, for $2 \le g \le 7$, are all real and interlace. Further automation (with the help of a symbolic computation software) of the comparison of the genus and center manifold expansions performed in \cite{bib:elt23a} will indicate whether this behavior continues for larger values of $g$. Using an alternate method to calculate the $z_g$'s, we were able to further verify this trend for $g$ up to 15. If this continues, understanding the structure of these polynomials, as well as the nature of the distribution of their zeros as $g \to \infty$, promises to be a rich and rewarding line of investigation.

\bigskip
\subsubsection{Higher order asymptotic expansions for map counts}
As mentioned in the introduction, the leading term ${\mathcal N}_{2 \nu,e}^\infty(g,j)$ of ${\mathcal N}_{2 \nu,e}(g,j)$ as the number of vertices $j$ goes to infinity is known \cite{bib:er11}. An analysis of expression \eqref{eq:eg_hyper} will provide a full asymptotic expansion for the map counts, of the form
\[
{\mathcal N}_{2 \nu,e}(g,j) = {\mathcal N}_{2 \nu,e}^\infty(g,j) \left(1 + \sum_{k=1}^p \frac{m_k}{j^k} + {\mathcal O} \left( \frac{1}{j^{p+1}}\right) \right), \quad j \to \infty,
\]
thereby clarifying the rate at which ${\mathcal N}_{2 \nu,e}(g,j)$ converges to ${\mathcal N}_{2 \nu,e}^\infty(g,j)$. Plots of the ratios ${\mathcal N}_{2 \nu,e}(g,j)/ {\mathcal N}_{2 \nu,e}^\infty(g,j)$ are provided in \cite{bib:elt23b} for $\nu = 2$ and $2 \le g \le 7$.

\bigskip
\subsubsection{Map counts and Bernoulli numbers} Since $e_g(1)$, the number of maps with zero vertices, is equal to zero, setting $\nu = 2$ and $z_0 = 1$ in \eqref{eq:eg} gives
\[
-C^{(g)} = \sum_{\ell=0}^{3g-3} q_\ell^{(0)}.
\]
An alternative, recursive, representation of $C^{(g)}$ was derived in \cite{bib:er14} and is given by
\begin{equation}
\label{eq:Cg}
\dfrac{-C^{(g)}}{2  (2g-3)!}  = \left[\frac{1}{(2g+2)!} - \frac{1}{(2g)! 12} + \frac{(1 - \delta_{2,g})}{(2g-1)!}\sum_{k=2}^{g-1} \frac{(2-2k)_{2g-2k+2}}{(2g-2k+2)!} C^{(k)}\right],
\end{equation}
with $C^{(2)}$ known from the corresponding expression of $e_2(z_0)$.
Using Equation \eqref{eq:Cg} and Maple \cite{bib:maple}, we realized that for $2 \leq g \leq 100$,
\[
-C^{(g)} = \dfrac{B_{2g}}{4g(g-1)},
\]
where the $B_{2g}$ are the even Bernoulli numbers. This in turn led to the following conjecture, mentioned in \cite{bib:elt23b}.
\begin{conj}
The orbifold Euler characteristic of the moduli space of genus $g$ Riemann surfaces, $\chi_{orb}(M_g)$, is related to the coefficients of the partial fraction expansion of the generating function $e_g(z_0)$ according to
\[
-C^{(g)} = \sum_{\ell=0}^{3g-3} q_\ell^{(0)} = \dfrac{B_{2g}}{4g(g-1)} = \chi_{orb}(M_g),
\]
where $M_g$ is the moduli space of Riemann surfaces of genus $g$.
\end{conj}
We recall that the relation
\[
\dfrac{B_{2g}}{4g(g-1)} = \chi_{orb}(M_g)
\]
was established by Harer and Zagier \cite{bib:hz86}, and that the Bernoulli numbers may be defined through the Gamma function according to
\[
\dfrac{\Gamma^\prime(x)}{\Gamma(x)} = \log x -\frac1{2x} - \sum_{g=1}^\infty \frac{B_{2g}}{2g} x^{-2g}.
\]
We are grateful to M.L. Yattselev for pointing out that Equation \eqref{eq:Cg} with $-C^{(g)} = {B_{2g}}/{4g(g-1)}$ is a special case of Gelfand’s reciprocity relation for Bernoulli numbers, thereby confirming the above conjecture. This relationship provides a novel method to calculate the orbifold Euler characteristic $\chi_{orb}(M_g)$ from the generating functions $e_g(z_0)$.

\end{document}